\documentclass[aip,preprint,12pt]{revtex4-1}
\usepackage{graphicx}
\usepackage{amsmath}
\usepackage{gensymb}
\begin{document}

\title{Tuning out disorder-induced localization in nanophotonic cavity arrays} 


\author{Sergei Sokolov}
\email[]{s.a.sokolov@uu.nl}
\homepage[]{http://www.nanolinx.nl/}
\affiliation{Nanophotonics, Debye Institute for Nanomaterials Science, Center for Extreme Matter and Emergent Phenomena, Utrecht University, P.O. Box 80.000, 3508 TA Utrecht, The Netherlands }
\affiliation{Complex Photonic Systems (COPS), MESA+ Institute for Nanotechnology, University of Twente, P.O. Box 217, 7500 AE Enschede, The Netherlands}
\author{Jin Lian}
\affiliation{Nanophotonics, Debye Institute for Nanomaterials Science, Center for Extreme Matter and Emergent Phenomena, Utrecht University, P.O. Box 80.000, 3508 TA Utrecht, The Netherlands }
\affiliation{Complex Photonic Systems (COPS), MESA+ Institute for Nanotechnology, University of Twente, P.O. Box 217, 7500 AE Enschede, The Netherlands}
\author{Emre Y\"uce}
\affiliation{Department of Physics, Middle East Technical Univeristy 06800, Ankara, Turkey}
\affiliation{Complex Photonic Systems (COPS), MESA+ Institute for Nanotechnology, University of Twente, P.O. Box 217, 7500 AE Enschede, The Netherlands}
\author{Sylvain Combri\'{e}}
\affiliation{Thales Research \& Technology, Route D\'{e}partementale 128, 91767 Palaiseau, France}
\author{Alfredo De Rossi}
\affiliation{Thales Research \& Technology, Route D\'{e}partementale 128, 91767 Palaiseau, France}
\author{Allard P. Mosk}
\affiliation{Nanophotonics, Debye Institute for Nanomaterials Science, Center for Extreme Matter and Emergent Phenomena, Utrecht University, P.O. Box 80.000, 3508 TA Utrecht, The Netherlands }
\affiliation{Complex Photonic Systems (COPS), MESA+ Institute for Nanotechnology, University of Twente, P.O. Box 217, 7500 AE Enschede, The Netherlands}

\begin{abstract}
\textbf{
Weakly coupled high-Q nanophotonic cavities are building blocks of slow-light waveguides and other nanophotonic devices. Their functionality critically depends on tuning as resonance frequencies should stay within the bandwidth of the device. Unavoidable disorder leads to random frequency shifts which cause localization of the light in single cavities. We present a new method to finely tune individual resonances of light in a system of coupled nanocavities. We use holographic laser-induced heating and address thermal crosstalk between nanocavities using a response matrix approach. As a main result we observe a simultaneous anticrossing of 3 nanophotonic resonances, which were initially split by disorder.
}
\end{abstract}
\date{\today}
\maketitle 

\section{introduction}
Optical circuits containing high-Q photonic crystal nanocavities have been proposed for delay lines, optical memory storage, optomechanics and quantum communication\cite{Yariv1999,Parra2007,Morichetti2012,Takesue2013,Kuramochi2014,Notomi2008, Fang2015,Tillmann2013,Gerace2009,Hamel2015}. However, unavoidable fabrication disorder in nanophotonic structures causes scattering which leads to frequency detuning, signal attenuation, and eventually localizes optical modes which ruins the transmission properties of the whole system\cite{Cooper2010,Sgrignuoli2015}. Even state-of-the-art nanofabrication with random spatial variations of only $\Delta x=1\ \rm{nm}$ can lead to resonance wavelength detunings of more than $\Delta \lambda =1\ \rm{nm}$ \cite{Taguchi2011, Ramunno2008}. 

Several methods have been proposed to tune nanocavities. Methods based on local refractive index change, such as photodarkening, photoactivation and photo-oxidation \cite{Faraon2008, Chen2011, Cai2013}, have limited tuning range and irreversibility as well as introduction of optical loss. Other methods which involve free-carriers or heat suffer from diffusion, namely free-carriers and heat diffuse far beyond the physical size of the cavity \cite{Notomi2005, Ruzicka2010, Sokolov2015}. Since in a system with a useful optical coupling between nanocavities elements should be placed physically close to each other, this results in unavoidable crosstalk in the tuning process. In addition, in high-$Q$ systems carriers dissipate on a time scale comparable to or shorter than the resonance lifetime.   

In this paper, we demonstrate a new approach for realignment of resonances of closely-spaced cavities based on holographic thermal tuning. By accurately measuring the thermal response of resonance wavelength in the system and obtaining the power settings required to hybridize cavities, we experimentally align an array of three nanocavities, which were initially misaligned by more than 200 unloaded resonator linewidths. Our method is scalable to large arrays and enables programmable photonic devices where circuit functionalities can be dynamically switched on and off.

\section{Sample and experimental setup}
\begin{figure}[h!]
 \centering \includegraphics[width=0.8\textwidth]{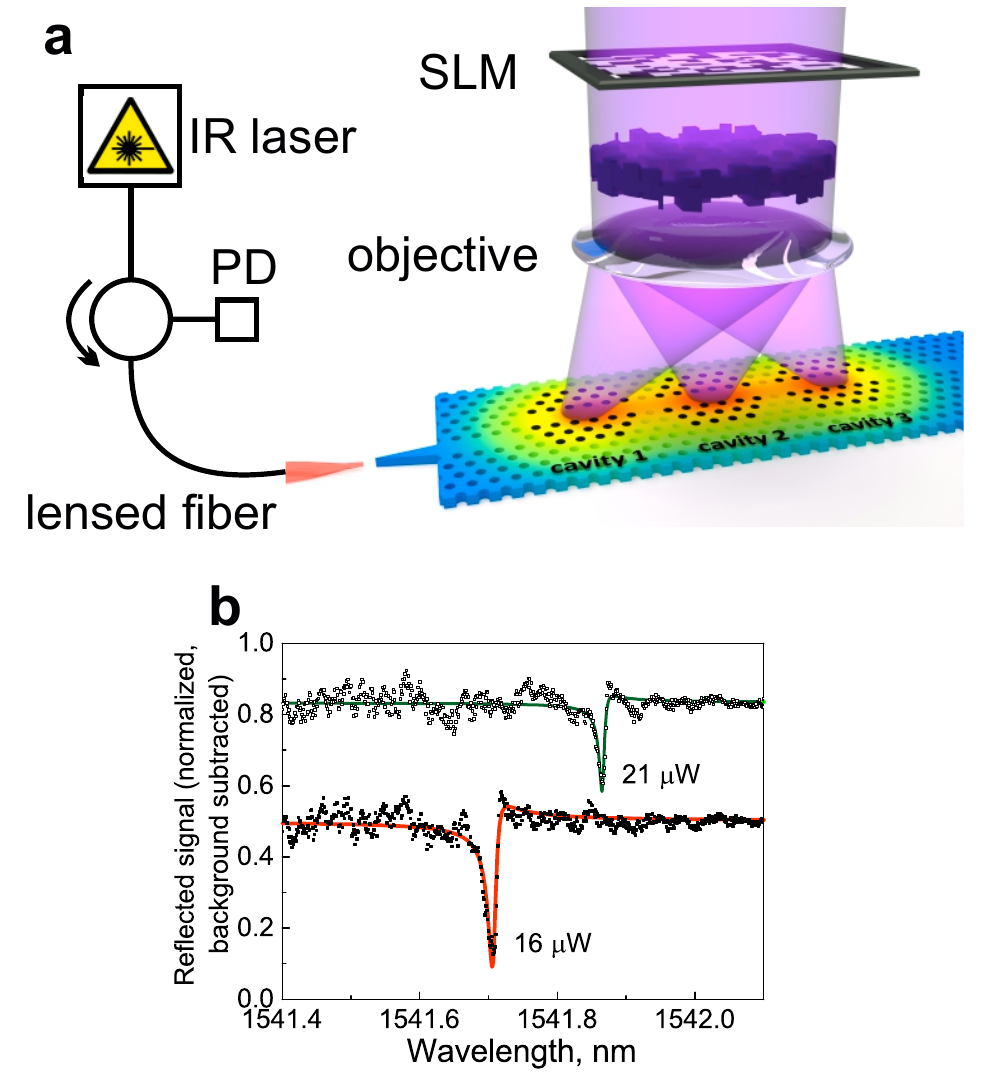}
 \caption{\label{fig1}\textbf{Schematic of the experiment.} \textbf{a} Pump light is  modulated by the SLM to holographycally create several focused spots on the sample. The SLM is imaged to the pupil of the objective. Continuous wave IR probe light from a tunable laser with TE polarization is coupled to the sample through a polarization maintaining lensed fiber and the reflected signal is collected on a photodiode using a fiber circulator. \textbf{b} Reflection spectra showing the resonance of cavity 3 for pump powers $16\ \mu \rm{W}$ and $21\ \mu \rm{W}$. Power applied to cavity 1 is $9\ \mu\rm{W}$. Solid lines represent Fano fits. The smoothed background is subtracted, offset is applied for clarity. }
\end{figure}
The sample under investigation is Ga$_{0.51}$In$_{0.49}$P air-suspended photonic crystal membrane with thickness $d=180 \rm{nm}$. Three nanocavities are defined in a photonic crystal waveguide with width of $0.98\sqrt{3}a$ made in a hexagonal lattice with period $a=485\ \rm{nm}$ and hole radius $r=0.28a$. They are defined by local width modulation\cite{Combrie2008} of the waveguide with a maximum hole shift of $0.0124a$. An out-of-line waveguide with a width $1.1\sqrt{3}a$ is used to couple light in and out of the system. The distance between cavity 1 and cavity 2 is $4.4\ \mu \rm{m}$, and the distance between cavity 2 and 3 is $4.9\ \mu \rm{m}$. Numerically calculated coupling rates are $\Gamma_{12}=0.00022\omega_0$ and $\Gamma_{23}=0.00015\omega_0$, where $\omega_0$ is the resonance wavelength of the cavity, $\Gamma_{jk}$ is a coupling rate between cavity j and k.

\begin{figure}[h]
 \centering \includegraphics[width=0.6\textwidth]{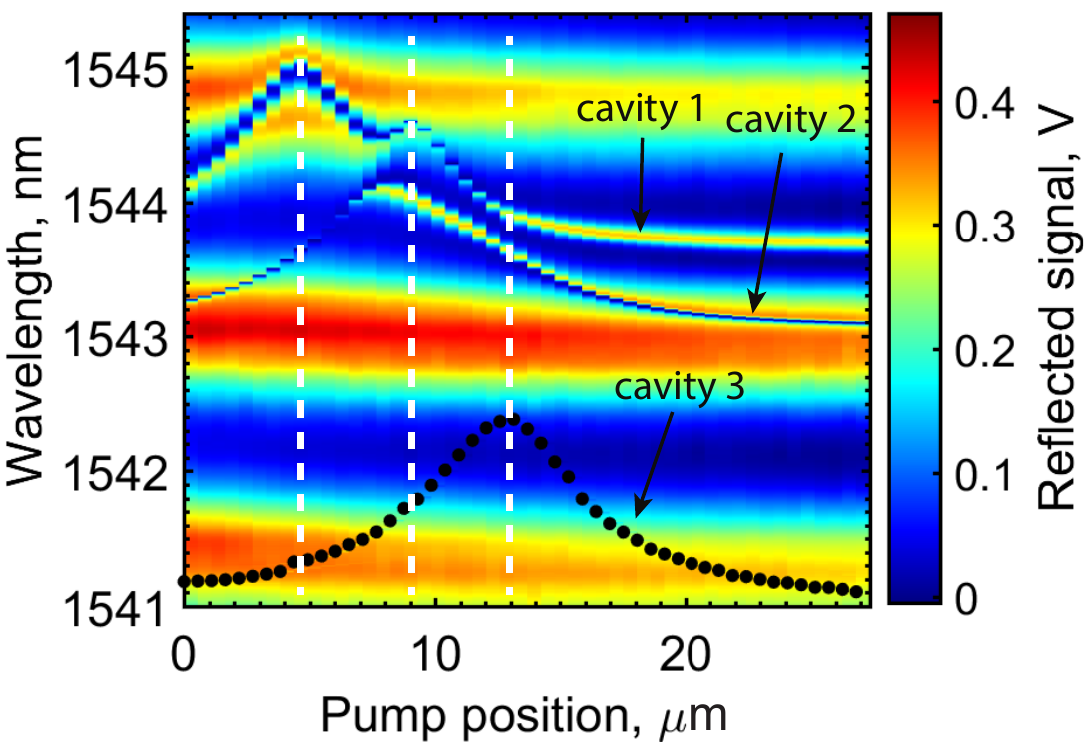}
 \caption{\label{linescan_fig}\textbf{Pump line-scan.}Response of cavity resonances for pump placed at different positions. The power on the surface of the sample is 32 $\mu\rm{W}$. Black dots indicate resonance of the cavity 3 for better visibility. White dashed lines indicate cavity positions.}
\end{figure}
The setup for the thermal control is shown in Fig. \ref{fig1}(a). CW pump light ($\lambda=405\ \rm{nm}$) is used to thermally tune nanocavities. A reflective liquid crystal spatial light phase modulator (SLM) was imaged onto the back aperture of the objective to holographically\cite{Pasienski2008} create several focused spots on the sample. Light was focused into tight spots with FWHM  $0.8\ \mu \rm{m}$. The absolute intensity of all pump spots was measured using a CCD camera placed in the reflection configuration and calibrated with a power meter. To attribute a particular resonance to a particular cavity we performed pump line-scans of the pump spot along the cavity array\cite{Sokolov2015,Lian2016}. The resulting linescan of our sample is presented in Fig. \ref{linescan_fig}. The closer the pump spot to the center of a particular cavity the bigger the overlap of the temperature profile created by the pump laser and mode profile of the cavity. This allows to easily identify resonances, as indicated in  Fig. \ref{linescan_fig}. The resonances of cavities 1 and 2 are easily visible in the graph. Resonance 3 is shallow and narrow (See Fig.2(b)), and while it is measured with signal-to-noise ratio >5 it is not obvious on the printed graph. So its position is indicated by black markers. In addition to resonance lines there are also wide fringes present in Fig. 2. These fringes results from interference in the system and are addressed in detail in Ref. \citenum{lian_arxiv_2016}. To extract resonance wavelengths and widths the reflection spectra were fitted with appropriate Fano line-shape of the resonances \cite{Zhou2014}, as it is shown in Fig. \ref{fig1}(b). 
\section{Thermal response matrix approach}
 When a photonic crystal membrane is heated with focused CW laser light the width of the temperature profile is more than 4 times larger than the diffraction-limited laser focus \cite{Chen2011, Sokolov2015}. Although this width can be tailored by properly selecting the material of the membrane and the ambient media\cite{Sokolov2015}, significant crosstalk remains for any choice of material. To account for that crosstalk we employ a response matrix approach. In the linear tuning regime, where the resonance shift is proportional to the applied power, we construct a matrix ($M_{ij}$) which expresses the response of the resonance wavelength ($\Delta\lambda_i$) to the applied pump powers $P_i$:        
   \begin{equation}\label{eq:mainequation}
  \Delta\lambda_i=\sum\limits_{j}M_{ij} \cdot P_j 
\end{equation}
 Diagonal elements of the response matrix determine the response of the addressed cavity and off-diagonal elements determine crosstalk to the neighbors. Ideally the response matrix values can be calculated \cite{Sokolov2015}, however they are sensitive to experimental details, therefore we use measured values. 

 \begin{figure}[h]
 \centering \includegraphics[width=0.9\textwidth]{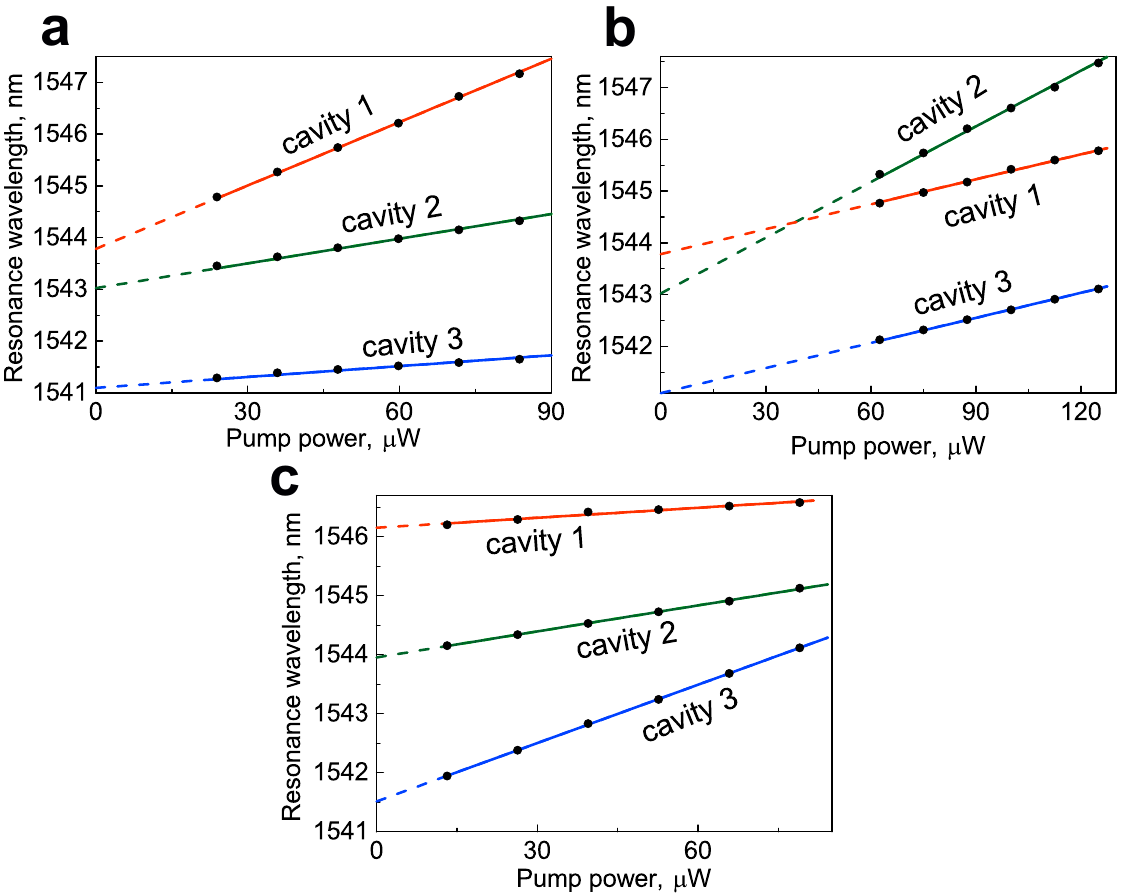}
 \caption{\label{fig2}\textbf{Determination of the thermal response matrix.} \textbf{a},\textbf{b},\textbf{c} Response curves of cavity resonances to pump spots placed on top of cavity 1 (\textbf{a}),2 (\textbf{b}),3 (\textbf{c}). Solid lines are line fits to experimental data and dashed lines are extrapolation of fitting curves to the zero power. In \textbf{c} cavity 1 was biased with $60\ \mu \rm{W}$ pump power to separate the resonances.}
 \end{figure}

We place a single pump spot on top of each cavity at a time and measure resonance redshift versus applied pump power. The result is presented in Figs. \ref{fig2}(a,b,c) for cavity 1, 2 and 3 respectively. In some cases when cavity resonances occur too close together, a second pump spot is placed to bias the resonance, as in Fig. \ref{fig2}(c). Resonance shifts are then fitted with linear dependencies, where slopes give thermal crosstalk values. The cavity directly under the pump spot displays the steepest slope whereas resonances of neighbor cavities have smaller slopes. The measured response matrix is
  \begin{equation}\label{eq:crosstalkmatrix}
 M=
 \begin{pmatrix}
 4.1 & 1.6 & 0.6 \\
 1.6 & 3.6 & 1.5 \\
 0.7 & 1.6 & 3.3 
 \end{pmatrix}
  \cdot 10^{-2} \frac{\rm{nm}}{\mu \rm{W}}
 \end{equation}
 Response values differ slightly from cavity to cavity as a consequence of positioning accuracy of pump spots, the difference in physical separation between cavities and possible differences in local surface reflectivity and thermal conductivity. On average first-neighbor crosstalk is 44\%, whereas second-neighbor crosstalk is 17\%. Such levels of thermal crosstalk cannot be neglected. We find bare wavelengths of nanocavity resonances to be $1543.78\pm0.11\ \rm{nm}$ for cavity 1, $1543.02\pm0.04\ \rm{nm}$ for cavity 2 and $1541.1\pm0.04\ \rm{nm}$ for cavity 3, which matches with reference measurements. Due to the disorder cavity 3 is detuned from cavities 1 and 2 by about $\Delta \lambda=2\ \rm{nm}$. The observed detunings can be explained by disorder, taking into account our fabrication accuracy of $\Delta x \approx 4\ \rm{nm}$. 
\section{Alignment of the array}
\begin{figure}[h]
 \centering\includegraphics[width=0.9\textwidth]{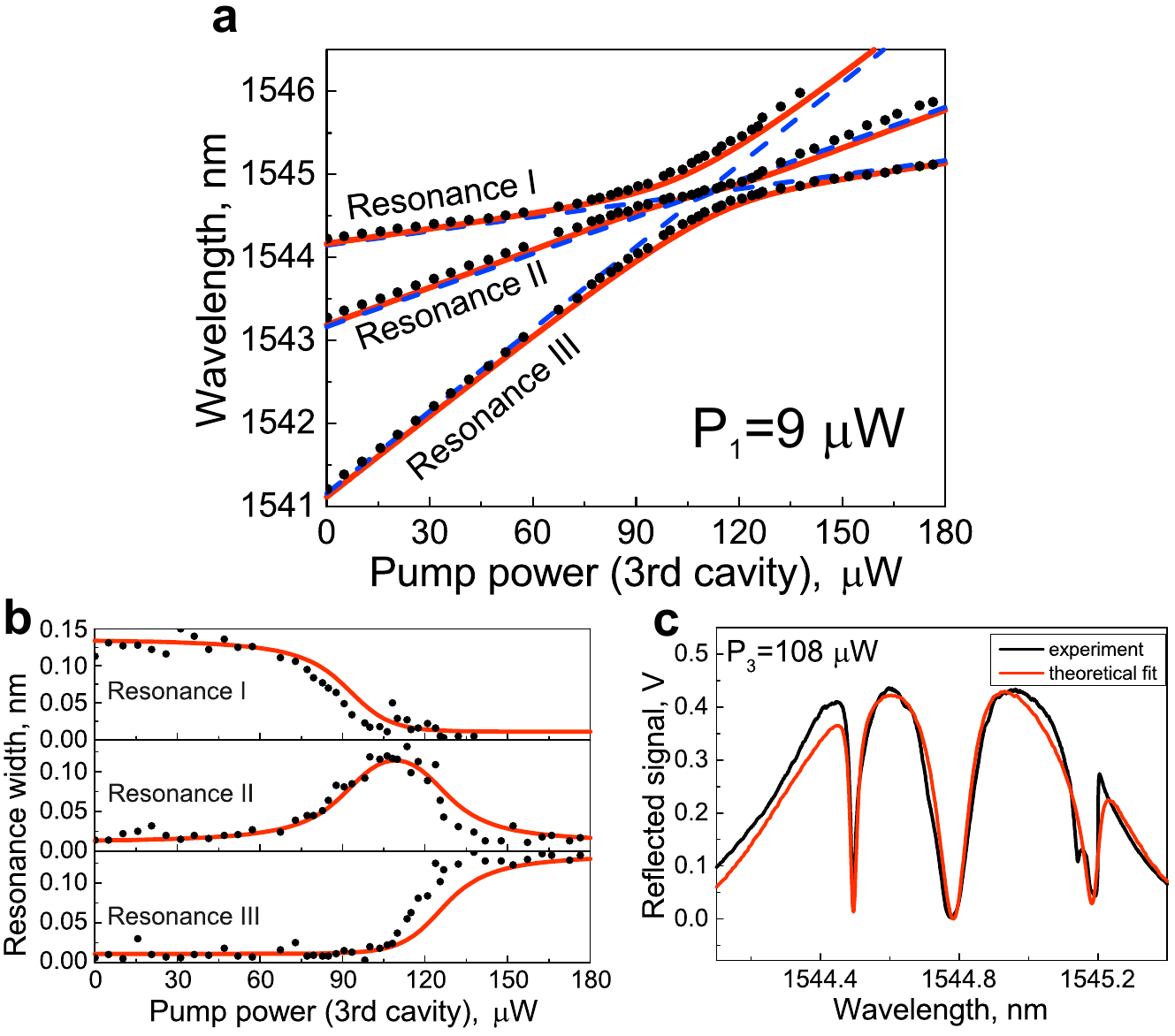}
 \caption{\label{fig3}\textbf{Alignment of cavity resonances.} Resonance positions \textbf{a} and widths \textbf{b} are obtained from reflection spectra by fitting Fano lineshapes. The pump power for cavity 1 was fixed to 9 $\mu$W, while the power for cavity 3 was increased from 0 to 180 $\mu$W. Red solid line is a fit by coupled-mode theory. Blue dashed lines represent uncoupled resonance wavelengths. \textbf{c} - Spectra of the sample for $P_3=108 \mu\rm{W}$ Red solid line represents Fano line fit with 3 resonance lines and 5th order polynomial as a background.}
 \end{figure}

 To find the required powers and the target wavelength at which all the resonances anticross we solve the resulting linear programming problem. The condition for this problem is that all applied powers are positive. We find that the target wavelength is $1544.79\ \rm{nm}$ and corresponding powers are $P_1=9\ \mu\rm{W}$, $P_2=0\ \mu\rm{W}$ and $P_3=108\ \mu \rm{W}$. As a result of the procedure the lowest applied pump power is always zero, therefore to align three resonances 2 pump spots are sufficient. In general to align $N$ cavities one needs to place at least $N-1$ pump spots on the surface of the sample, as in Eq. 1 we have $N$ equations but $N+1$ variables, therefore there is a freedom to select the power of one of the pump spots to be 0.  

In Fig. \ref{fig3}(a) we show the hybridization of the three resonances which is the main result of this paper. We set $P_1$ to the calculated value and gradually increase $P_3$ and plot resonance wavelengths. The resonances anticross and become fully hybridized at around $108\ \mu\rm{W}$. The reflection spectra for this power is shown in Fig. \ref{fig3}(c). In the region where resonances are hybridized they cannot be attributed to a particular cavity, hence we label them as resonance I, II and III. At low pump powers resonances I, II and III represent cavities 1,2 and 3 correspondingly. At high powers after the resonances anticross, resonance I localizes on cavity 3 and resonance III localizes on cavity 1. For pump powers higher than $140\ \mu\rm{W}$, resonance I becomes very weak and its wavelength cannot be extracted. 
 
In case of weak optical coupling between cavities the behavior of the resonances can be described by coupled-mode theory \cite{Haus1991} assuming only nearest-neighbor coupling. The result is presented in Fig. \ref{fig3}(a). To find coupling rates for our cavities the parameters of the coupled-mode model were adjusted in the region where cavities are coupled, i.e. for powers between 70 and 140 $\mu \rm{W}$. We find good agreement between theory and experiment and obtain coupling rates which are equal to $\Gamma_{12}=(0.90\pm 0.29)\cdot10^{-4}\cdot \omega_0$, $\Gamma_{23}=(1.92\pm0.27)\cdot10^{-4}\cdot \omega_0$, or equivalently $\Gamma_{12}=0.14\pm0.05$ nm and $\Gamma_{23}=0.30\pm0.04$ nm. 

The experimentally determined values of the coupling rates are different from the numerically calculated ones. This suggests that disorder affects not only the resonance wavelength and Q-factor, but also the mode profile of a cavity, which determines the coupling constant \cite{Haus1991}.

The widths of all three resonances are expected to change while the cavities anticross. Cavity 1 is the closest cavity to the input waveguide which suggests that it should have the largest width due to the leak to the waveguide. At the same time when cavities are coupled the leakage from second and third cavity to the waveguide should increase, and resonances II and III should broaden. In Fig. \ref{fig3}(b) the experimental dependence of the widths of the resonances on applied power $P_3$ is presented.  The width of the resonance I slowly decreases with pump power $P_3$ and becomes very narrow at high pump powers confirming that it has become localized on cavity 3. The inverse dependence is pronounced for resonance III which starts localized on cavity 3 and ends on cavity 1. At the point of anticrossing resonance II becomes the broadest.  
  \begin{figure}[h]
 \centering\includegraphics[width=0.7\textwidth]{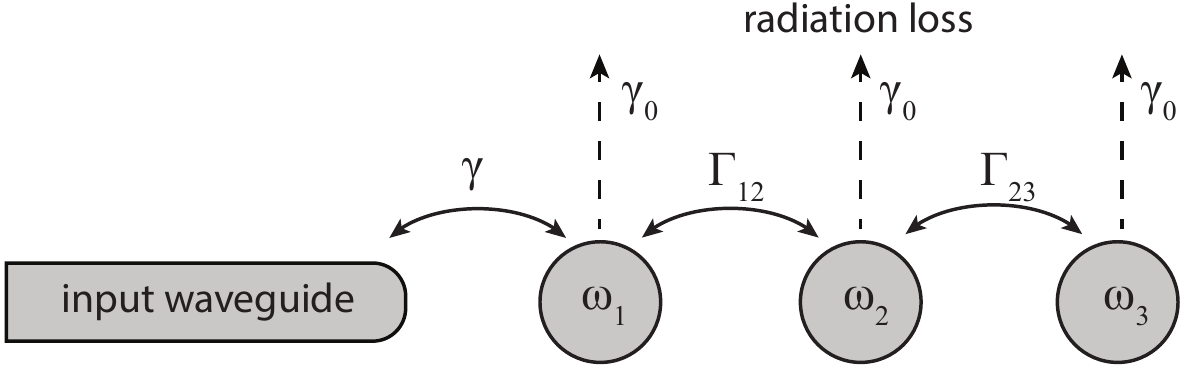}
 \caption{\label{fig_theory_scheme}\textbf{Model of the sample.} The light is coupled to the first cavity in the system with coupling rate $\gamma$, then each cavity in the array is coupled to the nearest neighbor by coupling constants $\Gamma_{12}$ and $\Gamma_{23}$.}
 \end{figure}
The width of the resonances can be predicted using temporal coupled-mode theory\cite{joannopoulos} where we take into account the influence of the input waveguide and radiation loss. The model is presented in Fig. \ref{fig_theory_scheme}. Cavities 2 and 3 are placed physically away from the input waveguide and therefore the coupling rates between them and the waveguide can be neglected. These resonances are visible in reflection only due to the light which leaks via the cavity 1. Only nearest neighbor coupling is taken into account. This result in the following system of equations:
\begin{equation}
 \begin{cases}
 S_-=-1+\sqrt{2\gamma}a_1,\\
 \frac{da_1}{dt}=-i\omega_{1}a_1-i\Gamma_{12}a_2-a_1\gamma-a_1\gamma_0+\sqrt{2\gamma},\\
 \frac{da_2}{dt}=-i\omega_{2}a_2-i\Gamma_{12}a_1-i\Gamma_{23}a_3-a_2\gamma_0,\\
 \frac{da_3}{dt}=-i\omega_{3}a_3-i\Gamma_{23}a_2-a_3\gamma_0\\
\end{cases}
 \end{equation}
Here $\gamma$ is a coupling rate to he input waveguide, $S_-$ is a complex amplitude of the outgoing waveguide mode, $a_1$, $a_2$ and $a_3$ are complex amplitude of the field inside cavities, $\omega_1$, $\omega_2$ and $\omega_3$ are bare frequencies of nanocavities and $\gamma_0$ determines intrinsic cavity decay rate.
We define two free parameters which are the loaded Q-factor of the first cavity ($Q_L$) and the intrinsic Q-factor ($Q_0$) of cavities assuming that it is the same for all cavities. We fit these free parameters outside the hybridization region, i.e. for power less than $70\ \mu W$ and more than 140 $\mu$W and obtain a good agreement with experiment (see Fig. \ref{fig3}(b)). The value for intrinsic Q-factor is found to be $Q_0=(1.42\pm0.45)\cdot10^5$, while the value of  loaded Q-factor for the first cavity is $Q_L=(1.12\pm0.05)\cdot10^4$. The numerically calculated value of the loaded Q-factor is 11000 which matches the value obtained from the experiment.

\section{Conclusion}
In conclusion, we have proposed and implemented a new method to program individual resonance wavelengths of coupled cavities. We corrected for thermal tuning crosstalk in the system to obtain independent tuning of resonances by measuring the thermal response matrix. We coupled three nanocavities and hybridized their local resonances into spatially extended modes. This shows that our method is capable of counteracting disorder in arrays of coupled cavities. The resonance wavelengths and widths were very well reproduced by a coupled-mode model. Our method can be easily extended to a larger number of nanocavities by adding more pump spots to the system. Resonance tuning with crosstalk compensation is likely to become a valuable method for any type of systems where disorder influences resonance states is required such as optomechanical systems \cite{Fang2015, Lauter2015}, many resonator systems \cite{Nozaki2014, Liapis2016, Gan2012, Mittal2014}, disordered necklace states \cite{Bertolotti2005, Sgrignuoli2015} and complex boson sampling networks \cite{ Spring2013, Crespi2013}.         
\section{Methods}
The experiment was performed in dry nitrogen atmosphere to minimize oxidation and surface water effects \cite{Chen2011}. The sample temperature was locked at $41.85\pm0.001\ \rm{\degree C}$. To further minimize these effects the illumination time of the surface with pump light was minimized to 270 ms during which the pumped measurement was taken. Nonetheless, minor surface effects were present in the data, resulting in resonance blueshifts and redshifts. To correct for that we performed reference measurement with no pump light immediately after each pumped measurement. Then surface effects for crosstalk matrix coefficients were eliminated by subtracting reference resonance values from pumped ones. As there is significant coupling between resonances 1 and 2  in the unpumped sample, to extract the bare resonance frequencies we used only data taken at pump powers >10 $\mu\rm{W}$. During the final measurement resonance shifts due to above mentioned effects were less than 80 pm for all presented datapoints. 
\section*{Funding}
European Research Council project (ERC) (279248), Nederlandse Organisatie voor Wetenschappelijk
Onderzoek (NWO).
\section*{Acknowledgments}
The authors would like to thank Sanli Faez, Henri Thyrrestrup and Willem Vos for helpful discussions and advises, Cornelis Harteveld for technical support. 
\bibliographystyle{naturemag}

\end{document}